\documentclass[prl,twocolumn,amsmath,amssymb,letterpaper,superscriptaddress,nofootinbib]{revtex4} 
\usepackage{graphicx}
\usepackage{amsmath}
\usepackage{amsfonts}
\usepackage{amssymb}
\usepackage[colorlinks,linkcolor=blue,filecolor=blue,urlcolor=blue,citecolor=blue]{hyperref}

\begin{document}

\title{Quantum Criticality of Liquid-Gas Transition in a Binary Bose Mixture}

\author{Li He}
\thanks{These authors contributed equally to this work}
\affiliation{College of Physics and Electronic Engineering, Shanxi University, Taiyuan 030006, China}
\author{Haowei Li}
\thanks{These authors contributed equally to this work}
\affiliation{CAS Key Laboratory of Quantum Information, University of Science and Technology of China, Hefei 230026, China}
\author{Wei Yi}
\affiliation{CAS Key Laboratory of Quantum Information, University of Science and Technology of China, Hefei 230026, China}
\affiliation{CAS Center For Excellence in Quantum Information and Quantum Physics, Hefei 230026, China}
\author{Zeng-Qiang Yu}
\email{zqyu.physics@outlook.com}
\affiliation{Institute of Theoretical Physics, Shanxi University, Taiyuan 030006, China}
\affiliation{State Key Laboratory of Quantum Optics and Quantum Optics Devices, Shanxi University, Taiyuan 030006, China}

\begin{abstract}
Quantum liquid, in the form of a self-bound droplet, is stabilized by a subtle balance between the mean-field contribution and quantum fluctuations.
While a liquid-gas transition is expected when such a balance is broken, it remains elusive whether liquid-gas critical points exist in the quantum regime.
Here we study the quantum criticality in a binary Bose mixture undergoing the liquid-gas transition.
We show that, beyond a narrow stability window of the self-bound liquid, a liquid-gas coexistence persists, which eventually transits into a homogeneous mixture. Importantly, we identify two distinct critical points where the liquid-gas coexistence terminates.
These critical points are characterized by rich critical behaviors in their vicinity, including divergent susceptibility, unique phonon-mode softening, and enhanced density correlations. The liquid-gas transition and the critical points can be readily explored in ultracold atoms confined to a box potential.
Our work highlights the thermodynamic approach as a powerful tool in revealing the quantum liquid-gas criticality, and paves the way for further studies of critical phenomena in quantum liquids.
\end{abstract}

\maketitle

\paragraph{Introduction.}
Liquid-gas transition is ubiquitous in nature, and serves as a paradigm of classical phase transitions. A well-known feature therein is the presence of critical points that mark the onset (or termination) of the liquid-gas coexistence~\cite{LandauBook}. In the quantum regime, exotic self-bound liquid states (dubbed quantum droplets) have recently been discovered in dipolar or binary Bose-Einstein condensates~\cite{Petrov2015,DipolarExp1,DipolarExp2,DipolarExp3,DipolarExp4,DipolarExp5, DropletExp1,DropletExp2,DropletExp3,DropletExp4,DropletExp6}, and experimental observations consistent with the liquid-gas coexistence have been reported in imbalanced mixtures~\cite{DropletExp2,DropletExp3,DropletExp4,DropletExp6}.
The discovery has stimulated extensive interest~\cite{review,DropletTheory2,santosdip1,santosdip2, blakiedip1,pfaudip,blakiedip2,DropletTheory3,DropletTheory4,DropletTheory5,DropletTheory6,DropletTheory7, DropletExpAdd1,DropletTheory8,DropletTheory9,UnequalMassTheory1,DropletExp5,DropletTheory10,DropletTheory11, UnequalMassTheory2,DropletTheory12,DropletTheory13,DropletTheory14,DropletTheory15,DropletTheory16,Hu3,DropletTheory17, shitaotheory,huhuitheory,FiniteTTheory1,FiniteTTheory2,FiniteTTheory3,FiniteTTheory4,FiniteTTheory6, DipMixTheory1,DipMixTheory2,Yin,Yin2,LHYgas1,LHYgas2},
culminating in the latest observation of dipolar supersolids in droplet crystals~\cite{supersolid1,supersolid2,supersolid3,supersolid4,supersolid5,supersolid6}.
However, little is known about the transition between the inhomogeneous liquid-gas coexistence
and the homogeneous liquid or gas phases, particularly in the thermodynamic limit.
A further important question is whether there exists a quantum analog of the critical point in experimentally relevant systems. Since quantum fluctuations play a key role in the formation of self-bound droplets~\cite{Petrov2015,santosdip1,santosdip2, blakiedip1,pfaudip}, they could lead to yet unexplored many-body phenomena at the critical points.
A systematic investigation of such quantum criticality would therefore offer further insight into quantum liquids and enrich our understanding of quantum phase transitions in general.

\begin{figure}[t!]
\includegraphics{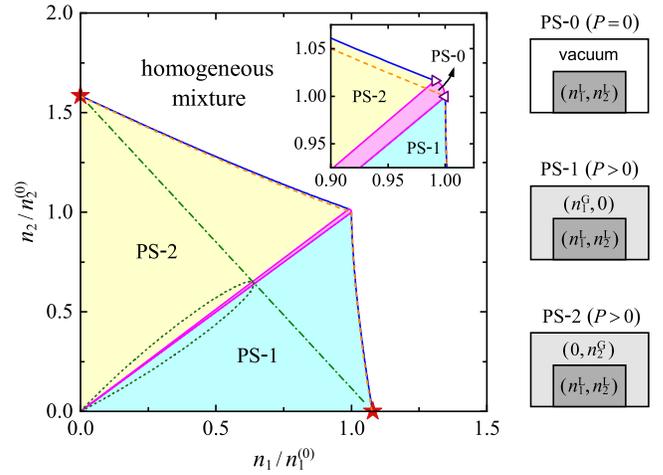}
\caption{Left: Ground-state phase diagram of a binary Bose mixture in the mean-field-unstable regime.
The liquid-gas separated states PS-1 and PS-2 terminate at the critical points denoted by $\bigstar$.
Blue solid lines: phase boundaries obtained by numerically solving the balance conditions.
Dashed lines: analytical boundaries given by Eq.~(\ref{Eq7}) and its counterpart with an exchange of species index.
The dash-dotted (dotted) line is the diffusive (mechanical) spinodal. Inset: enlarged view in the vicinity of $\big(n_1^{(0)}, n_2^{(0)}\big)$.
$\lhd$ ($\rhd$) denotes the point where the self-bound liquid reaches the evaporation threshold $\mu_1=0$ ($\mu_2=0$).
Right: illustrations of various inhomogeneous states with either zero or positive pressure. For all figures throughout this work, $\delta \tilde g = -0.08$, and $\lambda = 0.68$, which are relevant for spin mixtures of $^{39}$K atoms~\cite{DropletExp1,DropletExp2,DropletExp3}.} \label{Fig1}
\end{figure}

In this work, we address the questions above by studying the liquid-gas transition in a three dimensional binary Bose mixture using a general thermodynamic approach.
We find that, besides the self-bound state which is stable within a narrow window of densities~\cite{Petrov2015}, two types of liquid-gas coexistence generally exist, each with a distinct and fully polarized gas component (see Fig.~\ref{Fig1}).
Upon further tuning the densities, the system undergoes a transition from a liquid-gas separated state to a homogeneous phase.
Starting from the equation of state (EOS) with beyond-mean-field corrections,
we quantitatively characterize the phase diagram in the thermodynamic limit, and, crucially, reveal two critical points where the liquid-gas coexistence terminates.
Driven by density fluctuations, critical phenomena arise near the critical points, exemplified by the divergent susceptibility, the phonon-mode softening, and a dramatic enhancement of the correlation length.
Given the recent progress in trapping and probing cold atoms, both the transition and quantum criticality reported here can be readily investigated in a box potential~\cite{BoxTrap}.

\paragraph{Liquid-Gas Coexistence.}

We consider a three dimensional Bose mixture of cold atoms at zero temperature. 
The system features short-range interactions, with the interaction strengths $g_{ij}$ ($i,j=1,2$ labeling the atomic species). Here we consider interspecies attraction and intraspecies repulsion, with $g_{12}<0$ and $g_{11}, g_{22}>0$.
On the mean-field level, the system would collapse when $g_{12}<-g$, with $g\equiv\sqrt{g_{11}{g_{22}}}$. Such instability, however, can be dramatically modified once the quantum fluctuations are taken into account~\cite{Petrov2015}. We focus on the regime where $\delta \tilde g \equiv 1+g_{12}/g$ is very small. It follows that, for a homonuclear mixture with equal masses ($m_1=m_2=m$), the energy per volume can be written as~\cite{Petrov2015}
\begin{equation} \label{Eq1}
  \mathcal{E} = \sum_{i,j=1,2} \frac{g_{ij}}{2} n_i n_j
 + \frac{8m^{3/2}}{15\pi^2\hbar^3} \left(g_{11}n_1+g_{22}n_2\right)^{5/2},
\end{equation}
where $n_1$ and $n_2$ are the densities of the two species, respectively, 
$\hbar$ is the reduced Planck constant, and the second term represents the Lee-Huang-Yang corrections~\cite{LHYcorr}.

The EOS (\ref{Eq1}) is based on the presumption that the ground state is homogeneous. Yet, this is not true in the low-density limit under the mean-field instability.
For a concentration $n_1/n_2$ fixed at $\lambda\equiv\sqrt{\frac{g_{22}}{g_{11}}}$, the attractive and the repulsive mean-field contributions are mostly cancelled out, and the energy per particle reaches its minimum at the density~\cite{Petrov2015}
\begin{equation}  \label{Eq2}
  n_i^{(0)} = \frac{25\pi}{1024\, a^3} \sqrt{\frac{g}{g_{ii}}} \frac{\lambda^{5/2}}{(1+\lambda)^5} {\delta \tilde g}^2 ,
\end{equation}
with $a\equiv \sqrt{a_{11}a_{22}}$ ($a_{ij}=\frac{mg_{ij}}{4\pi\hbar^2}$ the \textit{s}-wave scattering length). As a result, when the total atom density fulfills $n<n^{(0)}$, a self-bound liquid state is formed. Here $n^{(0)}= n_1^{(0)}+n_2^{(0)}$. Such a state, referred to as PS-0 in Fig.~\ref{Fig1}, is stable even if the container of the system is removed, typical of the quantum droplet~\cite{DropletExp1,DropletExp3}.

The realization of the quantum droplet is not restricted to the exact density ratio $\lambda$. Thermodynamically, a stable self-bound liquid can be achieved under the conditions~\cite{DropletTheory5,DropletTheory10}
\begin{equation} \label{Eq3}
  P\left(n_1,n_2\right) = 0, \quad \mu_{1}\left(n_1,n_2\right)\leqslant0, \quad \mu_2\left(n_1,n_2\right)\leqslant0,
\end{equation}
where $P$ is the pressure, and $\mu_i$ is the chemical potential of species $i$. These conditions can be fulfilled within a narrow window of concentration, where the density of the self-bound liquid remains unchanged up to the order $\delta \tilde g^2$~\cite{Petrov2015}.

If the population of species $1$ increases further, such that the inequality $\mu_1\leqslant 0$ no longer holds, the PS-0 state will evolve into an inhomogeneous state with liquid-gas coexistence (PS-1 in Fig.~\ref{Fig1}). The balance conditions for the phase separation are
\begin{align}
  P\left(n_1^\textsc{l},n_2^\textsc{l}\right)  &= P\left(n_1^\textsc{g},0\right), \label{Eq4}\\
  \mu_1\left(n_1^\textsc{l},n_2^\textsc{l}\right) &= \mu_1\left(n_1^\textsc{g},0\right), \label{Eq5}\\
  \mu_2\left(n_1^\textsc{l},n_2^\textsc{l}\right) &< \mu_2\left(n_1^\textsc{g},0\right), \label{Eq6}
\end{align}
where $n_1^\textsc{l}$ and $n_2^\textsc{l}$ denote the densities of different species in the mixed liquid, and $n_1^\textsc{g}$ is the density of the coexisting gas of species 1.
While such liquid-gas coexistence has been numerically investigated in finite-size systems~\cite{DropletTheory11,DropletTheory12,DropletTheory13}, the phase transition between the phase-separated state and a homogeneous one is not yet well understood.

Indeed, the PS-1 state appears only at sufficiently low densities, and the ground state becomes a homogeneous liquid under the conditions $n_i=n_i^\textsc{l}$. The coexistence boundary, in terms of $n_1^\textsc{l}$ and $n_2^\textsc{l}$, can then be derived from Eqs.~(\ref{Eq4}) and (\ref{Eq5}) by eliminating $n_1^\textsc{g}$. Keeping densities to the leading order in $\delta \tilde g^2$, we obtain the analytical form of the phase boundary~\cite{SM}
\begin{align} \label{Eq7}
  3 \left(1+\lambda\right)^{5/2} (\tilde{n}_2^\textsc{l})^2 & - \left(\tilde{n}_1^\textsc{l} + \lambda \tilde{n}_2^\textsc{l} \right)^{3/2} \left[ (5+3\lambda)\tilde{n}_2^\textsc{l} - 2\tilde{n}_1^\textsc{l} \right] \nonumber \\
  & - 2\left(\tilde{n}_1^\textsc{l}  -\tilde{n}_2^\textsc{l} \right)^{5/2} = 0 \,,
\end{align}
where $\tilde{n}_i^\textsc{l} = n_i^\textsc{l} / n_i^{(0)}$.
As shown in Fig.~\ref{Fig1}, for small $\delta \tilde g$, the prediction of (\ref{Eq7}) is in good agreement with numerical calculations using Eqs.~(\ref{Eq4})(\ref{Eq5})(\ref{Eq6}).

By tuning the density ratio, one can also realize another kind of liquid-gas coexistence, the PS-2 state, where the gas phase consists only of atoms of species 2. Its phase boundary can be readily obtained by enforcing $\tilde n_1^\textsc{l}\leftrightarrow \tilde n_2^\textsc{l}$ and $\lambda\rightarrow \lambda^{-1}$ in Eq.~(\ref{Eq7}).

\begin{figure}[tbp]
\includegraphics{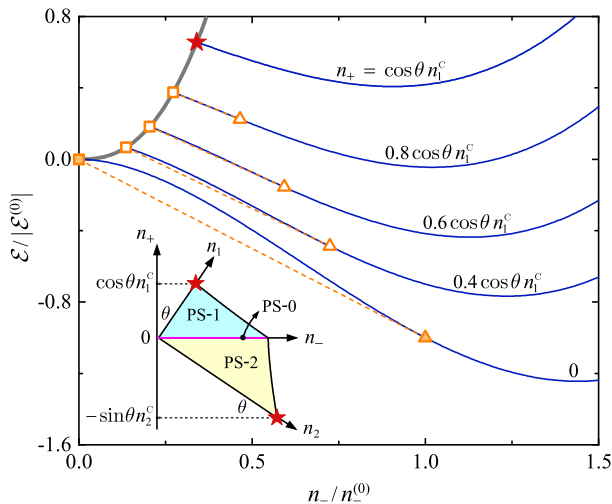}
\caption{EOS with the density variable $n_+$ fixed at different values. The dashed segments correspond to the energy of the inhomogeneous states. $\square$ and $\bigtriangleup$ denote the coexisting liquid and gas phase, respectively. The filled symbols highlight the case of $n_+=0$, where the liquid is self-bound. The gray bold line represents the EOS of a pure gas of species~1. For a better view, $\mathcal{E}$ is shifted by $\frac{1}{2}(g_{11}+g_{22})n_+^2$. $\mathcal{E}^{(0)}=\frac{1}{3} (g+g_{12}) n_1^{(0)}n_2^{(0)}$ is energy density of the self-bound liquid. The density variable $n_-$ is measured in the unit of $n_-^{(0)}\equiv n_1^{(0)}\sin\theta +n_2^{(0)}\cos\theta$.
Inset: phase diagram obtained in the single-mode approximation.
$\bigstar$ denotes the critical points.
} \label{Fig2}
\end{figure}

To shed more light on the phase-separated states, we introduce $n_+$ and $n_-$ to discern what we call the hard and soft degrees of freedom in response to the density variation~\cite{Petrov2015}
\begin{equation} \label{Eq8}
  \begin{pmatrix}
    n_+ \\ n_-
  \end{pmatrix}
  = \begin{pmatrix}
    \cos\theta & -\sin\theta \\ \sin\theta & \;\cos\theta
  \end{pmatrix}
  \begin{pmatrix}
    n_1 \\ n_2
  \end{pmatrix} ,
\end{equation}
with $\theta = \arctan \lambda$. A geometric interpretation of Eq.~(\ref{Eq8}) can be clearly seen from the inset of Fig.~\ref{Fig2}. For a given $n_+$, the allowed values of $n_-$ must be greater than the physical bound $n_-^\textrm{min}$, where the system becomes a single-species gas. In the low-density regime, since
$ \left| \frac{\partial \mu_i}{\partial n_+} \right|\gg \left|\frac{\partial \mu_i}{\partial n_-}\right|$,
the thermodynamic balance requires the hard-mode variable $n_+$ to be almost invariant in the coexisting phase (hence the name hard mode), enabling a single-mode approximation. As shown in Fig.~\ref{Fig2},  when $n_+$ lies within an appropriate range (expression given later), $\mathcal{E}(n_-)$ changes from concave to convex in the starting segment, meaning the energy of the phase-separated state (dashed lines) is lower than that of the homogenous state.
Under the tangent Maxwell construction, the coexistence condition is thus
\begin{equation} \label{Eq9}
   \left(n_-^\textsc{l}-n_-^\textrm{min}\right) \left. \frac{\partial\mathcal{E}}{\partial n_-}\right|_{n_-^\textsc{l}} = \mathcal{E}(n_+,n_-^\textsc{l}) - \mathcal{E}(n_+,n_-^\textrm{min}) \,.
\end{equation}
For positive (negative) $n_+$, Eq.~(\ref{Eq9}) gives the boundary of the PS-1 (PS-2) state, consistent with Eq.~(\ref{Eq7});
for $n_+=0$, it recovers the zero-pressure condition for the self-bound liquid.

Using Eqs.~(\ref{Eq4}) and (\ref{Eq5}), we have checked that the variation of $n_+$ in the coexisting liquid and gas phases vanishes at the order $\delta\tilde g^2$, which represents the accuracy of the single-mode approximation~\cite{SM}.
For a relatively larger $|\delta \tilde g|$, the terms neglected in the Lee-Huang-Yang contribution in Eq.~(\ref{Eq1}) would result in a higher-order shift of the phase boundary.

\paragraph{Quantum Criticality.}

In part of the coexistence region, the homogeneous phase appears as a metastable state, similar to the superheated liquid in the classical liquid-gas transition~\cite{LandauBook}. When densities fall below the diffusive spinodal line fixed by $\gamma_1\gamma_2=\gamma_{12}^2$ ($\gamma_i\equiv\frac{\partial \mu_i}{\partial n_i}$ and $\gamma_{12}\equiv \frac{\partial \mu_1}{\partial n_2}$), a homogeneous mixture becomes unstable against local density fluctuations.
Note that the mechanical spinodal line, along which the compressibility diverges, lies inside the unstable region (see Fig.~\ref{Fig1}).

The diffusive spinodal line can be derived using the EOS (\ref{Eq1}), and corresponds to a straight line in the $n_1$-$n_2$ plane, satisfying $n_1/n_1^\textsc{c}+n_2/n_2^\textsc{c}=1$~\cite{SM}. Since the spinodal must be enveloped by the coexistence boundaries, the difference between the separated phases vanishes at the densities $(n_1^\textsc{c},0)$ or $(0,n_2^\textsc{c})$. In other words, the liquid-gas transitions terminate at these critical points. To the order $\delta \tilde g^2$, we find
\begin{equation} \label{Eq11}
  n_i^\textsc{c} = \frac{16}{25} \left(1+\frac{g}{g_{ii}}\right) n_i^{(0)}.
\end{equation}
In the representation of $(n_+,n_-)$, the liquid-gas coexistence only occurs within the interval $-\sin\theta n_2^\textsc{c} <n_+< \cos\theta n_1^\textsc{c}$, while the homogeneous ground state evolves smoothly at either larger or smaller $n_+$, reminiscent of the supercritical regime of a classical liquid-gas transition.

Importantly, in the vicinity of these critical points, density fluctuations dominate and give rise to abundant critical behaviors.
Thermodynamically, the quantum criticality is manifested in the singular behavior of the susceptibilities $\chi^0_{ij}\equiv (\partial n_i/\partial \mu_j)_{\mu_{3-j}}$, which characterize the static response to density perturbations. With some algebra, $\chi^0_{ij}$ can be rewritten as
\begin{align}
  \chi_{ii}^0 = \frac{\gamma_{3-i}}{\gamma_1\gamma_2-\gamma_{12}^2} , \qquad
  \chi_{12}^0 =\chi_{21}^0 = \frac{-\gamma_{12}}{\gamma_1\gamma_2-\gamma_{12}^2} ,
\end{align}
which become divergent at either critical point.

\begin{figure}[tbp]
\includegraphics{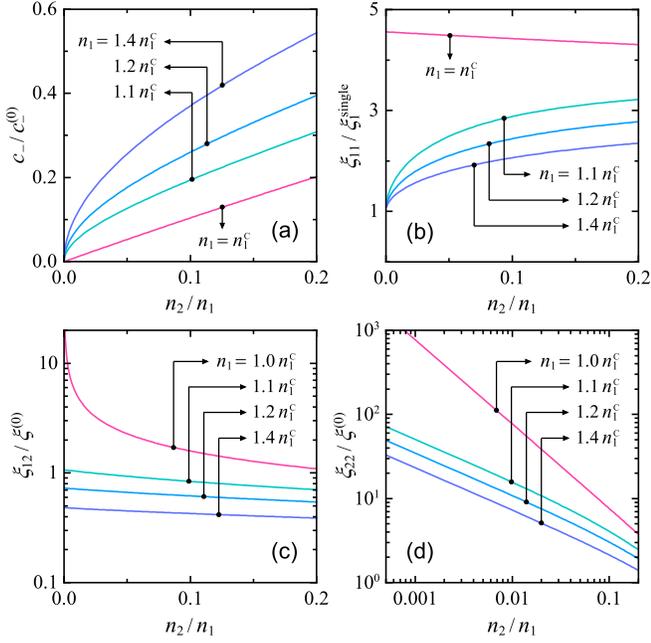}
\caption{(a) Sound velocity $c_-$ and (b)-(d) correlation lengths $\xi_{ij}$ as functions of the minority concentration $n_2/n_1$, for different $n_1$.
} \label{Fig3}
\end{figure}

Another related critical phenomenon is the softening of the phonon excitations.
Specifically, we derive the sound velocities of the phonon modes using the standard hydrodynamic approach~\cite{SM}
\begin{equation}
  c_\pm = \sqrt{ \frac{1}{2m} \Big[\gamma_1 n_1 + \gamma_2 n_2 \pm \sqrt{ ( \gamma_1 n_1 - \gamma_2 n_2 )^2 + 4 \gamma_{12}^2 n_1n_2} \Big]},
\end{equation}
where $c_-$ vanishes at either critical point.
At first glance, this seems quite natural, since only one phonon mode can survive as the density of the minority species approaches zero.
However, it is only at the critical points that
$c_-$ exhibits a unique linear dependence on the vanishing minority density.
For instance, in the low-concentration limit with $n_1=n_1^\textsc{c}$, the sound velocity $c_- = \frac{5\sqrt{\lambda}}{2\sqrt{2}} c_-^{(0)} \tilde n_2$,
where $\tilde n_2=n_2/n_2^{(0)}$, and $ c_-^{(0)} = 4\sqrt{gn^{(0)}\sqrt{n^{(0)} a^3}/5\sqrt{\pi}m} $ is the sound velocity of the self-bound liquid~\cite{SM}.
In contrast, we find $c_-\propto \sqrt{\tilde n_2}$ in the low-concentration limit with a fixed $n_1>n_1^\textsc{c}$. Such distinction [see Fig.\;\ref{Fig3}(a)] provides a clear signature for detecting the critical points.
Note that, for $n_1<n_1^\textsc{c}$, $c_-$ becomes imaginary in the spinodal region, indicating a dynamic instability.

The quantum criticality is also manifested in the dramatic changes in the correlation length.
The relative probability of finding two particles of a given species at distance $r$ is measured by the pair-distribution function $\mathcal{D}_{ij}(r)$, which at large separation takes the form $\mathcal{D}_{ij}(r\rightarrow \infty) = 1- \frac{\xi_{ij}}{\sqrt{2n_in_j} \pi^2 r^4}$, with $\xi_{ij}$ the correlation length~\cite{LHY1957,FeenbergBook}. Thus, the combined length scale $(\xi_{ij}/\sqrt{n_in_j})^{1/4}$ represents a characteristic distance, over which $\mathcal{D}_{ij}$ deviates considerably from unity. To determine $\xi_{ij}$, we employ the hydrodynamic approach to derive the dynamic density response function, which is connected to the Fourier transform of $\mathcal{D}_{ij}$ through the fluctuation-dissipation theorem. $\xi_{ij}$ is then extracted from the asymptotic expansion of $\mathcal{D}_{ij}$~\cite{SM}.

At a critical point, for instance $n_1 = n_1^{\textsc{c}}$ and $n_2\rightarrow 0$, $\xi_{ij}$ behaves like~\cite{SM}
\begin{gather}
  \xi_{11}  \rightarrow \xi_1^\textrm{single}  \Big( 1+ \frac{1}{\sqrt{|\delta \tilde g|}}\Big),
\\
  \xi_{12}  \sim \frac{\xi^{(0)}}{\sqrt{\tilde n_2}}\rightarrow  \infty \,,
  \qquad
  \xi_{22}  \sim \frac{\xi^{(0)}}{\tilde n_2}\rightarrow  \infty \,,
\end{gather}
where $\xi_1^\textrm{single}=\left.\hbar/\sqrt{2mn_1\gamma_1}\right|_{n_2=0}$ is the healing length of a single-species Bose gas~\cite{LHY1957,LevSandroBook}, $\xi^{(0)}=\frac{\sqrt{3}\hbar(\sqrt{g_{11}}+\sqrt{g_{22}})}{g\sqrt{2m|\delta \tilde g| n^{(0)}}}$ is the typical surface thickness of a self-bound droplet~\cite{Petrov2015}.
By contrast, for the case with any given $n_1>n_1^{\textsc{c}}$ (and $n_2\rightarrow 0$), we have~\cite{SM}
\begin{align}
  \xi_{11} \rightarrow \xi_1^\mathrm{single} ,
  \;\quad
  \xi_{12} \rightarrow  A\xi^{(0)}  ,
  \;\quad
  \xi_{22} \sim \frac{\xi^{(0)}}{\sqrt{\tilde n_2}}\rightarrow \infty \,,
\end{align}
with the coefficient $A$ given in Supplemental Material.
The distinction between these two situations, as illustrated in Figs.~\ref{Fig3}(b)-3(d), reflects the significant enhancement of density correlations at the quantum criticality, and can be tested experimentally using the Bragg spectroscopy.
Note that the discussions above apply to the other critical point by exchanging the species labels.

It is worth noting that the structures of the PS-1 and PS-2 states resemble that of the partially miscible states recently predicted for $g_{12}>g$~\cite{Petrov2021}, where the Lee-Huang-Yang correction also plays a crucial role in establishing the coexistence equilibrium. The key difference is that, under the repulsive interspecies interactions in Ref.~\cite{Petrov2021}, the phase separation occurs between gaseous phases and the dynamic instability is due to the out-of-phase fluctuations of two species; whereas in our case ($g_{12}<-g$), the underlying quantum criticality originates from the in-phase fluctuations.

\begin{figure}[tbp]
\includegraphics{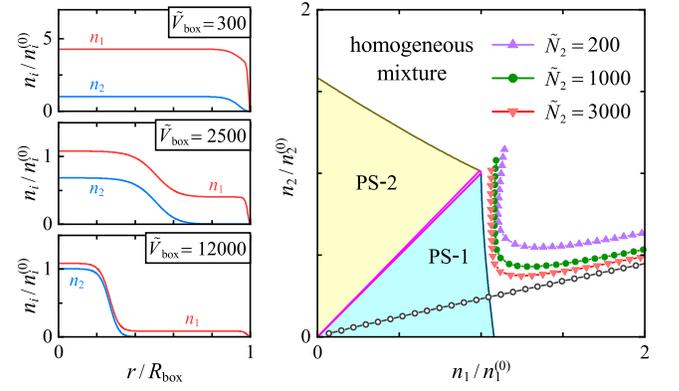}
\caption{Left: atomic density profiles in a box trap of different sizes with $(\tilde N_1, \tilde N_2)=(5000,1000)$.
Right: trajectory of the central density during an adiabatic expansion with various atom numbers. For comparison, the result for the case with $\delta \tilde g=0$ and $\tilde N_2=3000$ is also shown ($\circ$). All the simulations are performed with the concentration $\tilde N_1/\tilde N_2=5$. } \label{Fig4}
\end{figure}

\paragraph{Density Profiles.}

The predicted phase diagram can be experimentally verified with ultracold atoms confined in a box potential~\cite{BoxTrap}. We numerically simulate the atomic density profiles by solving the extended Gross-Pitaevskii equation~\cite{Petrov2015}  
\begin{align}
  \left[ - \frac{\hbar^2\nabla^2}{2m} + \frac{\partial }{\partial n_i} \mathcal{E}\big[n_1(\mathbf{r}),n_2(\mathbf{r})\big] -  \bar\mu_i^\textrm{box} \right] \psi_i(\mathbf{r}) = 0\,,
\end{align}
where $\psi_i$ is the condensate wavefunction satisfying the boundary condition $\psi_i =0$ at hard walls of the box, and $\bar\mu_i^\textrm{box}$ is the global chemical potential fixed by the normalization condition $\int d\mathbf{r} \left|\psi_i(\mathbf{r})\right|^2 = N_i$. For simplicity, we consider a spherical box trap of radius $R_\textrm{box}$, and introduce dimensionless variables $\tilde N_i=N_i/n_i^{(0)}{\xi^{(0)}}^3$ and $\tilde V_{\rm box}=\big(R_\textrm{box}/ \xi^{(0)}\big)^3$. The ground state of the system is numerically determined through imaginary-time evolutions under the split-step method~\cite{splitstep}.

As illustrated in the left column of Fig.~\ref{Fig4},
for sufficiently small $\tilde V_\textrm{box}$, the atomic densities are almost uniform, except for a thin layer close to the boundary.
When $\tilde V_\textrm{box}$ increases beyond a certain threshold, the liquid-gas coexistence appears, and the density profile exhibits a shell structure, with a liquid core immersed in a single-component gas of the majority species.
Here the densities of the two coexisting phases roughly obey the relation $\tilde n_1^\textsc{g}=\tilde n_1^\textsc{l}-\tilde n_2^\textsc{l}$ with $\tilde n_{i}^\textsc{g}=n_{i}^\textsc{g}/n_i^{(0)}$, consistent with results from the single-mode approximation.
As $\tilde V_\textrm{box}$ further increases, the gas in the outer shell becomes extremely dilute, and the liquid core is essentially self-bound with the densities approaching the saturated values $(n_1^{(0)},n_2^{(0)})$.
These results imply that, under the liquid-gas coexistence, a clear distinction between the gas and the liquid phases can be observed during an adiabatic expansion---while the outer gas shell diffuses throughout the box, the liquid core retains a finite volume.

Further, the phase diagram can be readily extracted from the flattop density profile. The right panel of Fig.~\ref{Fig4} shows the evolution of the central density when $\tilde V_\textrm{box}$ gradually increases at a fixed concentration $\tilde N_1/\tilde N_2=5$. As the phase separation sets in, the density trajectory turns upward abruptly and follows the phase boundary of the PS-1 state. Such a behavior is in stark contrast to the case without a liquid-gas transition (empty circles). By choosing a concentration $\tilde N_1/\tilde N_2\ll 1$, the boundary of PS-2 state can also be obtained.

Due to the finite-size effect, the phase boundary constructed in this way shows some deviations from that in the thermodynamic limit. The deviation becomes less pronounced when $\tilde N_i\gtrsim 1000$ (see Fig.~\ref{Fig4}).
Under our choice of parameters (see Fig.~\ref{Fig1}), the condition $(\tilde N_1, \tilde N_2)=(1,1)$ corresponds to $(N_1,N_2)=(1.16,1.71)\times 10^3$, which means that experiments with atom numbers of the order  $10^6$ should suffice.

\paragraph{Discussion.}

Adopting a thermodynamic approach, we quantitatively characterize the liquid-gas coexistence in a mean-field-unstable Bose mixture, and reveal the underlying quantum criticality.
The liquid-gas transition considered here also occurs in heteronuclear mixtures such as $^{41}$K-$^{87}$Rb~\cite{DropletExp4} and $^{23}$Na-$^{87}$Rb~\cite{DropletExp6},
where similar phase diagrams can be established using the analytical EOS therein~\cite{Petrov2021,SM}.

For future studies, it is desirable to explore the quantum liquid-gas criticality in lower dimensions~\cite{DropletTheory2,DropletTheory9,DropletTheory14,DropletTheory15,DropletTheory16,DropletTheory17}, with three-body~\cite{ThreeBodyRef1,ThreeBodyRef2} or dipolar~\cite{DipolarExp1,DipolarExp2,DipolarExp3,DipolarExp4,DipolarExp5, santosdip1,blakiedip1,santosdip2,blakiedip2,DipMixTheory1,DipMixTheory2, supersolid1,supersolid2,supersolid3,supersolid4,supersolid5,supersolid6} interactions, or at finite temperatures~\cite{FiniteTTheory1,FiniteTTheory2,FiniteTTheory3,FiniteTTheory4,FiniteTTheory6}.
At finite temperatures in particular, the interplay between quantum and thermal fluctuations may affect the nature of the condensation~\cite{FiniteTTheory1,FiniteTTheory6,FiniteTTheory7}, giving rise to intriguing critical behaviors.


\begin{acknowledgments}
We thank Lan Yin and Shizhong Zhang for helpful discussions. This research is supported by NSFC under Grants No.~12174230 and 12147215 (Z.-Q.Y.), Grant No. 12104275 (L.H.), and Grant No.~11974331 (W.Y.); and partially by the Fund for Shanxi 1331 Project of Key Subjects Construction.
\end{acknowledgments}

\paragraph{Notes Added.}
Recently, we became aware of a related work~\cite{XLCui}, where the liquid-gas transition and the associated critical behavior are discussed in a different setup.

\end{document}


\centerline{\textbf{Supplemental Material for}}

\title{Quantum Criticality of Liquid-Gas Transition in a Binary Bose Mixture}

\author{Li He}
\affiliation{College of Physics and Electronic Engineering, Shanxi University, Taiyuan 030006, China}
\author{Haowei Li}
\affiliation{CAS Key Laboratory of Quantum Information, University of Science and Technology of China, Hefei 230026, China}
\author{Wei Yi}
\affiliation{CAS Key Laboratory of Quantum Information, University of Science and Technology of China, Hefei 230026, China}
\affiliation{CAS Center For Excellence in Quantum Information and Quantum Physics, Hefei 230026, China}
\author{Zeng-Qiang Yu}
\affiliation{Institute of Theoretical Physics, Shanxi University, Taiyuan 030006, China}
\affiliation{State Key Laboratory of Quantum Optics and Quantum Optics Devices, Shanxi University, Taiyuan 030006, China}

\maketitle

\renewcommand{\theequation}{S\arabic{equation}}
\renewcommand{\thefigure}{S\arabic{figure}}

In this Supplemental Material, we provide details for the determination of the phase boundaries, the hydrodynamic approach in deriving the density response functions, and the characterization of critical behaviors.

\section{I. \quad  Phase Diagram}

\subsection{1.1 \quad Phase boundaries of Liquid-Gas Coexsitence}

In the PS-1 state defined in the main text, the system consists of a quantum-liquid component, coexisting with a gas of species 1.
We determine the phase boundary of the PS-1 state from the balance conditions
\begin{align}
 \mu_1\left(n_1^\textsc{l},n_2^\textsc{l}\right) &= \mu_1\left(n_1^\textsc{g},0\right) \, , \label{S1}\\
 P\left(n_1^\textsc{l},n_2^\textsc{l}\right)  &= P\left(n_1^\textsc{g},0\right)  \, ,\label{S2}
\end{align}
where, as defined in the main text, $n_1^\textsc{l}$ and $n_2^\textsc{l}$ are the densities of different species in the mixed liquid, and $n_1^\textsc{g}$ is the density of the coexisting gas.

From Eq.~(8) in the main text, one can reexpress the equilibrium densities using $(n_+^\textsc{l},n_-^\textsc{l})$ and $(n_+^\textsc{g},n_-^\textsc{g})$.
It follows from the equation of state (EOS) that, in the low-density regime with $n_ia^3 \lesssim\delta\tilde g^2$, we have $\left| \frac{\partial \mu_i}{\partial n_+} \right|\gg \left|\frac{\partial \mu_i}{\partial n_-}\right|$. Here $a$ is defined in Eq.~(2) of the main text.
Therefore, to meet the requirement of Eq.~(\ref{S1}), the hard-mode variable $n_+$ differs but little in the two coexisting phases.
Back in the $(n_1, n_2)$ representation, we have
\begin{align}
  n_1^\textsc{g} = n_1^\textsc{l} - \lambda n_2^\textsc{l} - \Delta, \label{S3}
\end{align}
where $\Delta$ is small and of the same order as $n_+^\textsc{l}-n_+^\textsc{g}$. As will be self-consistently checked later, the densities of coexisting phases are at most of the order $\delta \tilde g^2$, and $\Delta $ is at most of the order $\delta\tilde g^3$.
We then substitute Eq.~(\ref{S3}) into Eqs.~(\ref{S1}) and (\ref{S2}), and keep terms to ensure that the mean-field and the Lee-Huang-Yang contributions are of the same order.
It follows that
\begin{align}
  g_{11}\Delta + \delta g n_2^\textsc{l} + \frac{4m^{3/2}}{3\pi^2\hbar^3}g_{11} \left[ \left(g_{11}n_1^\textsc{l}+g_{22}n_2^\textsc{l}\right)^{3/2} - \left(g_{11}n_1^\textsc{l}-gn_2^\textsc{l}\right)^{3/2} \right] = 0 \, ,\label{S4}
  \\
  \left(g_{11}n_1^\textsc{l}- gn_2^\textsc{l}\right)\Delta + \delta g n_1^\textsc{l} n_2^\textsc{l} + \frac{4m^{3/2}}{5\pi^2\hbar^3}  \left[ \left(g_{11}n_1^\textsc{l}+g_{22}n_2^\textsc{l}\right)^{5/2} - \left(g_{11}n_1^\textsc{l}-gn_2^\textsc{l}\right)^{5/2} \right]  = 0 \, ,\label{S5}
\end{align}
where $\delta g = g_{12}+g$.
By eliminating $\Delta$, we obtain an analytical expression for the coexistence boundary
\begin{align} \label{S6}
  \lambda |\delta \tilde g| {n_2^\textsc{l}}^2 - \frac{32a^{3/2}}{15\sqrt{\pi}} \left[ \left(\frac{n_1^\textsc{l}}{\lambda}+\lambda n_2^\textsc{l}\right)^{3/2} \! \left( 5n_2^\textsc{l}+3\lambda n_2^\textsc{l} -2\frac{n_1^\textsc{l}}{\lambda} \right) + 2\left(\frac{n_1^\textsc{l}}{\lambda}-n_2^\textsc{l}\right)^{5/2}\right] = 0 \, .
\end{align}
It is then straightforward to recast Eq.~(\ref{S6}) into the dimensionless form given in the main text [Eq.~(7) therein], by scaling $n_i^\textsc{l}$ with $n_i^{(0)}$.

We note that, at the specific density ratio $n_1/n_2 = \lambda$, Eq.~(\ref{S6}) reproduces the solution of self-bound state, $n_i^\textsc{l}=n_i^{(0)}$.
When $n_2^\textsc{l}\rightarrow 0$, the coexistence line given by Eq.~(\ref{S6}) terminates at the critical point $(n_1^\textsc{c},0)$. The phase boundary of the PS-2 state can be obtained following a similar procedure, by switching the species labels. As illustrated in Fig.~1 of the main text, phase boundaries from the analytic expressions above agree well with those from numerical calculations.
To plot the analytic coexistence line, it is more convenient to rewrite Eq.~(\ref{S6}) as
\begin{align}\label{S7n}
  n_1^\textsc{l} \,=\, \frac{225\pi}{1024a^3} \frac{\lambda^3 \:\! y^4}{\left[(1+\lambda y)^{3/2}(5y+3\lambda y-2)+2(1-y)^{5/2}\right]^2}\delta\tilde g^2 \,,
\end{align}
where $y\equiv \lambda n_2^\textsc{l}/n_1^\textsc{l}$. Then, by varying $y$ in the interval of $0$ to $1$, the boundary of the PS-1 state is obtained. The same technique is also used to plot the PS-2 boundary.
Note that, according to Eq.~(\ref{S7n}), $n_i^\textsc{l}\sim \delta\tilde g^2$, which is consistent with our choice of truncations leading to Eqs.~(\ref{S4}) and (\ref{S5}).

\begin{figure}[tbp]
\includegraphics{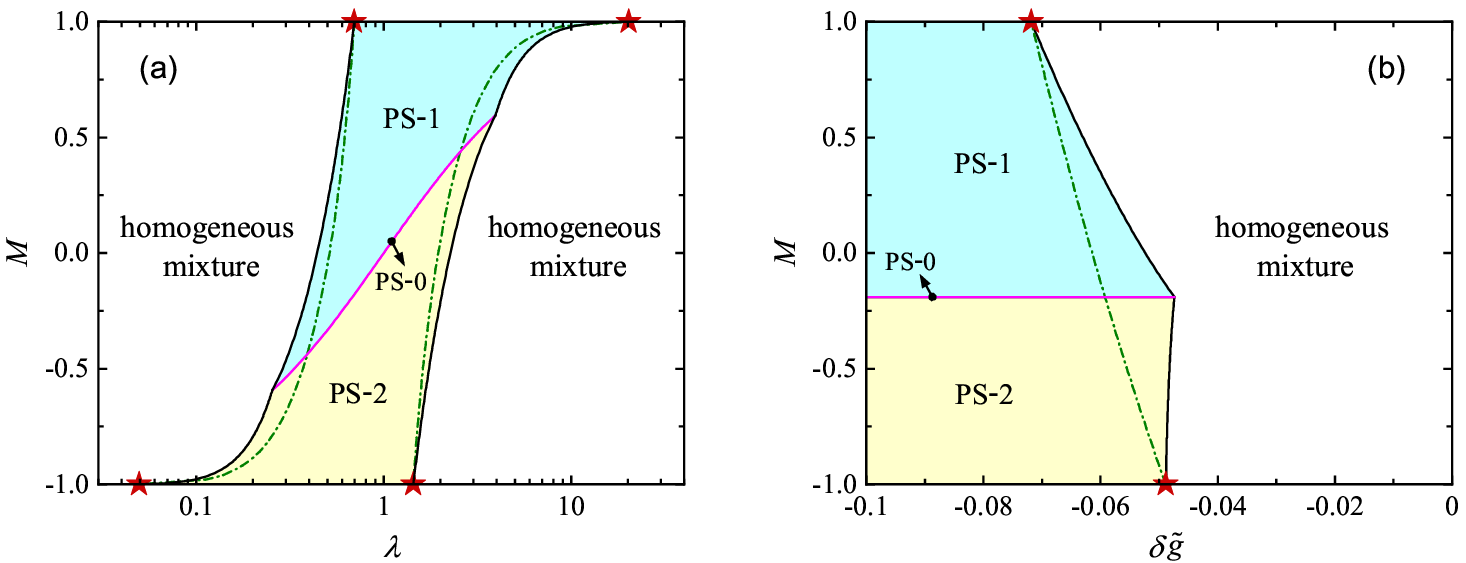}
\caption{ Phase diagram of a Bose-Bose mixture in terms of different variables. Black solid lines: the liquid-gas coexistence boundaries given by Eq.~(\ref{S6}) and its counterpart with a switch of species index. Dash-dotted lines: the diffusive spinodal given by Eq.~(\ref{spinodal}). $\bigstar$ denotes the critical points. Parameters: in (a), $na^3=0.002\delta \tilde g^2$; in (b), $na^3=10^{-5}$ and $\lambda=0.68$.
} \label{FigS1}
\end{figure}

It is worth mentioning that the transition from the liquid-gas coexistence to the homogeneous phase can be also triggered by tuning the interaction parameters $\lambda$ or $\delta\tilde g$. In Fig.~\ref{FigS1}, we show the phase diagrams in the $\lambda$-$M$ plane and in the $\delta\tilde g$-$M$ plane, respectively, based on the analytical results derived above. Here $M\equiv(N_1-N_2)/(N_1+N_2)$ is the polarization of the mixture. Note that, in the relevant experimental setup, such as the spin mixture of $^{39}$K atoms~\cite{DropletExp1,DropletExp3}, when one of the interaction parameters $\{a,\lambda,\delta\tilde g\}$ varies, the other two change correspondingly. In this sense, the most convenient way to experimentally probe the phase diagram is to adjust the atomic densities with fixed interaction parameters.
This is the scheme we discussed in the main text.

\subsection{1.2 \quad Single-Mode Approximation}

Since $n_+$ barely changes in the two coexisting phases,
we can simplify calculations and gain further insight by taking the single-mode approximation, assuming
\begin{align}
  n_{+}^\textsc{l} = n_{+}^\textsc{g} \, . \label{S7}
\end{align}
Accordingly, we can determine the equilibrium structure from the single-variable EOS $\mathcal{E}(n_-)$, with a given $n_+$. By applying the tangent Maxwell construction~\cite{HuangBook}, the densities of the coexisting phases are fixed by
\begin{align}
  \left(n_-^\textsc{l}-n_-^\textsc{g}\right) \left. \frac{\partial\mathcal{E}}{\partial n_-}\right|_{n_-^\textsc{l}} = \mathcal{E}(n_-^\textsc{l}) - \mathcal{E}(n_-^\textsc{g}) \,. \label{S8}
\end{align}
Combining this result with Eq.~(\ref{S7}) then yields the coexistence condition (7) in the main text.

In the PS-1 state, the coexisting gas phase only consists of atoms of species 1, hence Eqs.~(\ref{S7}) and (\ref{S8}) can be rewritten as
\begin{gather}
  n_1^\textsc{g} = n_1^\textsc{l} - \lambda n_2^\textsc{l} \, , \label{S9}  \\
  n_2^\textsc{l} \left[ \lambda \mu_1(n_1^\textsc{l},n_2^\textsc{l}) + \mu_2(n_1^\textsc{l},n_2^\textsc{l})\right] = \mathcal{E}(n_1^\textsc{l},n_2^\textsc{l}) - \mathcal{E} (n_1^\textsc{g},0) \,. \label{S10}
\end{gather}
In deriving Eq.~(\ref{S10}), we have used $ \left.\frac{\partial\mathcal{E}}{\partial n_-}\right|_{n_-^\textsc{l}}= \sin\theta \mu_1 (n_1^\textsc{l},n_2^\textsc{l})+ \cos\theta\mu_2(n_1^\textsc{l},n_2^\textsc{l}) $.

Note that Eqs.~(\ref{S9}) and (\ref{S10}) are not exactly equivalent to the balance conditions Eqs.~(\ref{S1}) and (\ref{S2}).
But the subtle difference between them does not affect the phase boundary, as long as the densities are kept to leading orders in $\delta \tilde g^2$. Specifically, by using (\ref{S9}) to eliminate $n_1^\textsc{g}$ in (\ref{S10}), we reproduce Eq.~(\ref{S6}) for the phase boundary.
To understand the equivalence of these derivations,
we notice that, according to Eqs.~(\ref{S4}) and (\ref{S5}), the leading correction to the difference $n_+^\textsc{l}-n_+^\textsc{g}$ is of the order $\delta \tilde g^3$. Therefore, fixing $n_+$ in the EOS would result in an inaccuracy of at most the same order in determining the phase boundaries.  On the other hand, in the tangent Maxwell condition Eq.~(\ref{S8}), the two parts of the interaction energy exhibit different
density scalings: the mean-field terms are mostly cancelled out and their net contribution is proportional to $\delta \tilde g n^2$; while the Lee-Huang-Yang terms give a contribution proportional to $n^{5/2}$.
As a result, the coexistence line predicted by the single-mode theory only involves densities of the order $\delta \tilde g^2$, consistent with the approximation used in the preceding subsection.


\subsection{1.3 \quad Spinodal}

In part of phase-separated region, enclosed by the coexistence line and diffusive spinodal, the homogeneous mixture can persist as a metastable phase. The diffusive spinodal is fixed by the condition $\gamma_1\gamma_2=\gamma_{12}^2$, where $\gamma_i$ and $\gamma_{12}$ can be readily derived from the EOS with the following results
\begin{align}
  \gamma_1 \, &= \,  \frac{\partial^2 \mathcal{E}}{\partial n_1^2} \, = \, g_{11} \left( 1 + \frac{16}{\sqrt{\pi}} \frac{a^{3/2}}{\lambda} \sqrt{\frac{n_1}{\lambda} + \lambda n_2} \right),  \label{gamma1}
  \\
  \gamma_2 \, &= \,  \frac{\partial^2 \mathcal{E}}{\partial n_2^2} \, = \, g_{22} \left( 1 + \frac{16}{\sqrt{\pi}} \lambda a^{3/2} \sqrt{\frac{n_1}{\lambda} + \lambda n_2} \right),  \label{gamma2}
  \\
  \gamma_{12} \, &= \, \frac{\partial^2 \mathcal{E}}{\partial n_1\partial n_2} \,= \, g \left( -1 + \delta \tilde g +  \frac{16}{\sqrt{\pi}} a^{3/2} \sqrt{\frac{n_1}{\lambda} + \lambda n_2} \right).   \label{gamma12}
\end{align}
Thus, the spinodal equation can be explicitly written as
\begin{align} \label{spinodal}
  \sqrt{\frac{n_1}{\lambda} + \lambda n_2} = \frac{\sqrt{\pi}}{8a^{3/2}} \frac{\lambda}{(1+\lambda)^2} |\delta \tilde g| \,,
\end{align}
where we have omitted the terms at higher orders in $\delta \tilde g$.

For fixed interaction parameters, Eq.~(\ref{spinodal}) describes a straight line in the $n_1$-$n_2$ plane. Since the diffusive spinodal meets the liquid-gas coexistence boundaries at the critical points $(n_1^\textsc{c},0)$ and $(0,n_2^\textsc{c})$, it can be rewritten in the form $n_1/n_1^\textsc{c}+n_2/n_2^\textsc{c}=1$, as given in the main text.

In Fig.~\ref{FigS1}, we also show the variation of the spinodal lines with the interaction parameters $\lambda$ and $\delta \tilde g$.

\subsection{1.4 \quad Case with Unequal Masses}

In the above calculations, we assume the two species are of equal masses. Now we turn to heteronuclear mixtures with unequal masses. For sufficiently small  $|\delta g|$, the EOS of the system is governed by $\mathcal{E}= \frac12 \sum_{i,j} g_{ij}n_in_j + \mathcal{E}_\textsc{lhy}$, where the Lee-Huang-Yang energy density $\mathcal{E}_\textsc{lhy}$ can be written as~\cite{Petrov2021}
\begin{align}
  \mathcal{E}_\textsc{lhy} \, &= \,  \frac{8 m_1^{3/2}} {15\pi^2\hbar^3} (g_{11}n_1)^{5/2} \mathcal{I}\!\left(\tfrac{m_2}{m_1},\tfrac{g_{22}n_2}{g_{11}n_1}\right) \,,
\end{align}
with the dimensionless function $\mathcal{I}(z,x)$ given by
\begin{align} \label{I_fun}
  \mathcal{I}(z,x) \,=  \;  \frac{1}{2(z^2-1)} & \left[ \left(x^2z^2+ 2z^2-7xz-2\right) \sqrt{\frac{x}{z}+1} - \left(xz^3-3z^2+6xz+2\right) \frac{\sqrt{1+xz}}{\sqrt{z^2-1}}\, \mathbb{F}\!\left(\arccos\tfrac1z, \tfrac{xz}{1+xz}\right) \right.
  \nonumber \\
  & \left. + \left(2x^2z^4+7xz^3-3x^2z^2-3z^2+7xz+2\right) \frac{\sqrt{1+xz}}{\sqrt{z^2-1}}\, \mathbb{E}\!\left(\arccos\tfrac1z, \tfrac{xz}{1+xz}\right) \right] .
\end{align}
Here $\mathbb{F}(\varphi,\upsilon)$ is the elliptic integral of the first kind, and $\mathbb{E}(\varphi,\upsilon)$ is the elliptic integral of the second kind.


The liquid-gas coexistence boundaries can be analytically derived within the single-mode approximation.
By substituting the EOS into Eq.~(\ref{S10}), we find the following expression for the transition line of the PS-1 state after some straightforward algebra
\begin{align} \label{PS1}
  \lambda |\delta \tilde g| {n_2^\textsc{l}}^2 \,= \, \frac{4m_1^{3/2}g^{3/2}}{15\pi^2\hbar^3}
  \left[ 5\bigg(\frac{n_1^\textsc{l}}{\lambda}\bigg)^{\!3/2\!}
  \bigg( n_2^\textsc{l}-\frac{n_1^\textsc{l}}{\lambda} \bigg) \mathcal{J}\!\left(\tfrac{m_2}{m_1},\tfrac{g_{22}n_2}{g_{11}n_1}\right) + 3\bigg(\frac{n_1^\textsc{l}}{\lambda}\bigg)^{\!5/2} \mathcal{I}\!\left(\tfrac{m_2}{m_1},\tfrac{g_{22}n_2}{g_{11}n_1}\right) + 2\bigg(\frac{n_1^\textsc{l}}{\lambda}-n_2^\textsc{l}\bigg)^{\!5/2} \right] .
\end{align}
Here $ \mathcal{J}(z,x) \, \equiv \, \mathcal{I}(z,x) - \frac{2}{5} x\mathcal{I}'(z,x) $, with $\mathcal{I}'(z,x)$ being the shorthand notation for $\frac{\partial}{\partial x} \mathcal{I}(z,x) $. At the specific density ratio $n_1/n_2=\lambda$, Eq.~(\ref{PS1}) gives the densities of the self-bound liquid of unequal masses
\begin{align} \label{n_sb}
  n_i^{(0)} = \frac{25\pi}{1024a^3} \sqrt{\frac{g}{g_{ii}}} \left(\frac{m_2}{m_1}\right)^{3/2}\! \frac{\lambda^{5/2}}{ \left[\mathcal{I}\!\left(\frac{m_2}{m_1},\lambda\right)\right]^2} \delta \tilde g^2 \,.
\end{align}
In the limit $n_2^\textsc{l}\rightarrow 0$, the coexistence line terminates at the critical point $(n_1^\textsc{c},0)$, with $n_1^\textsc{c}$ given by
\begin{align} \label{n1c}
  n_1^\textsc{c} \,=\, \frac{\pi}{64a_{11}^3} \frac{\delta \tilde g^2}{\left[ 1+\frac45\lambda\,\mathcal{I}'(z,0) + \frac{4}{15} \lambda^{2\;} \!\mathcal{I}''(z,0) \right]^2 } \,,
\end{align}
where $\mathcal{I}''(z,x)$ is the shorthand for $\frac{\partial^2}{\partial x^2} \mathcal{I}(z,x)$.
The functions $\mathcal{I}'(z,0)$ and $\mathcal{I}''(z,0)$ in Eq.~(\ref{n1c}) can be explicitly written in compact forms
\begin{align}
  \mathcal{I}'(z,0) \,=\,  \frac{15\,z}{4\left(z^2-1\right)} \Big(\frac{z^2}{\sqrt{z^2-1}} \arccos \tfrac1z -1\Big)
  \,, \qquad
  \mathcal{I}''(z,0) \,=\, \frac{45\,z^2}{16\,(z^2-1)} \Big(\frac{z^2-2}{\sqrt{z^2-1}} \arccos \tfrac1z +1\Big) \,.
\end{align}
In the case of $z=1$,
\begin{align}
  \mathcal{I}(1,x) = (1+x)^{5/2} \,, \qquad  \mathcal{J}(1,x) = (1+x)^{3/2} \,, \qquad
  \mathcal{I}'(1,0) = \frac{5}{2} \,, \qquad \mathcal{I}''(1,0) = \frac{15}{4} \,,
\end{align}
accordingly, Eqs.~(\ref{PS1}), (\ref{n_sb}) and (\ref{n1c}) recover the results for the equal-mass case given in the main text. Through a simple switch of species index, the expression for the boundary of the PS-2 state can also be obtained.

\begin{figure}[tbp]
\includegraphics{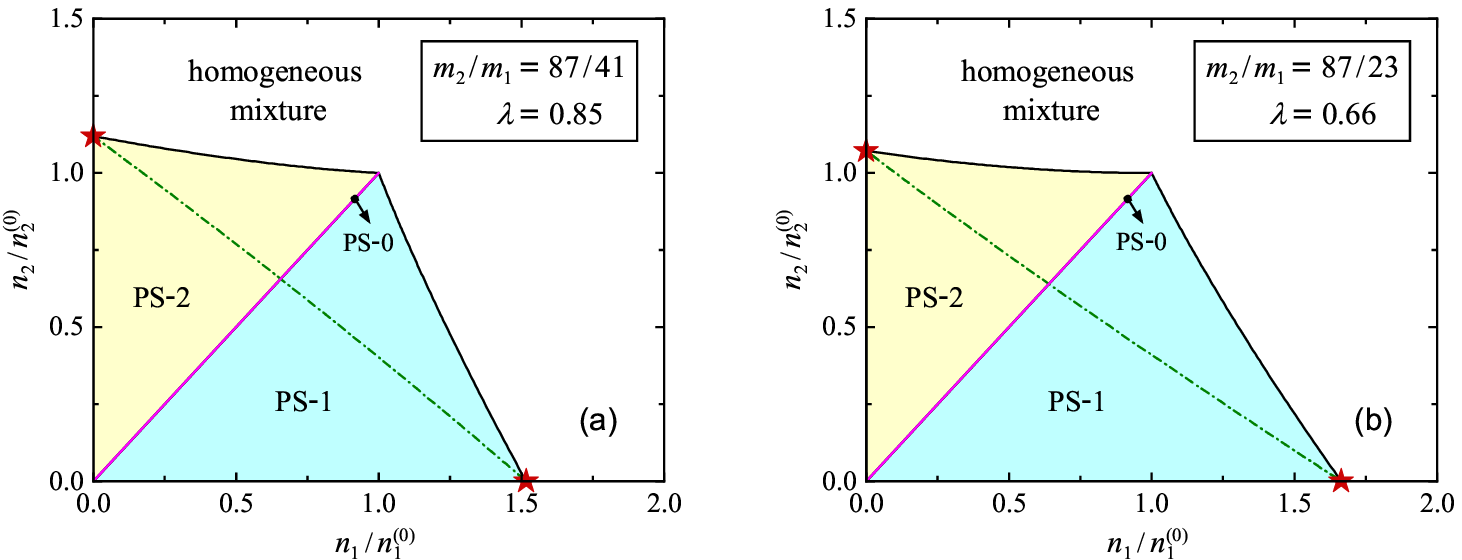}
\caption{  Phase diagram predicted for (a) $^{41}$K-$^{87}$Rb mixture and for (b) $^{23}$Na-$^{87}$Rb mixture within the single-mode approximation. The boundaries of the liquid-gas coexistence (black solid lines) meet the diffusive spinodal (dash-dotted line) at the critical points denoted by $\bigstar$. The interaction parameter $\lambda$ is set to the values reported in Refs.~\cite{DropletExp4} and \cite{DropletExp6}.
} \label{FigS2}
\end{figure}

In Fig.~\ref{FigS2}, we show the phase diagram predicted by the single-mode theory, using realistic parameters for the $^{41}$K-$^{87}$Rb mixture and $^{23}$Na-$^{87}$Rb mixture, respectively.
The overall landscapes of the phase diagrams are qualitatively similar to that of the equal-mass case.
The diffusive spinodal, which can be directly determined from the EOS, is also plotted in this figure. Generally, under the mass imbalance, it is no longer a straight line in the $n_1$-$n_2$ plane, although its curvature is typically small under typical experimental parameters.

\section{II. \quad Hydrodynamic Approach}

In this section, we employ the hydrodynamic approach~\cite{PethickBook,LevSandroBook} to study the density fluctuations in response to a weak, time-dependent perturbation, which are useful for identifying the critical phenomena. We consider the perturbative Hamiltonian

\begin{align}
  \hat H_\textrm{pert} = -\sum_{i=1,2} \alpha_i \hat\rho_{\mathbf{q},i}^\dag e^{i\omega t+\eta t} + \textrm{H.c.} \,,
\end{align}
where $\hat\rho_{\mathbf{q},i}^\dag$ is the density-fluctuation operator of species $i$ with $\mathbf{q}$ the wave vector,  $\alpha_i$ is the perturbation strength, and $\omega$ is the frequency. The positive infinitesimal $\eta$ ensures that the system is governed by the unperturbed Hamiltonian at $t=-\infty$. Below, we focus on the equal-mass case. The generalization to the unequal-mass case is straightforward.

\subsection{2.1 \quad Density Response Functions}

Assuming the perturbation varies slowly in space, we write the energy of the system as a functional of the local density $n_i({\bf r},t)$
\begin{align}
  E = \int \mathrm{d}\mathbf{r} \left[ \sum_{i=1,2} \frac{\hbar^2}{2m} (\nabla \sqrt{n_i})^2  + \mathcal{E}(n_1,n_2) - \sum_{i=1,2} \alpha_i \left( n_i e^{i(\mathbf{q\cdot r}-\omega t)}+ \textrm{c.c.}\right)e^{\eta t} \right] . \label{S15}
\end{align}
Here we have omitted the kinetic energy term of the super flow, since it does not affect the density fluctuations in the linear response regime.

The evolution of the condensate wavefunction $\psi_i(\mathbf{r}, t) = \sqrt{n_i(\mathbf{r}, t)} \, e^{i\phi_i(\mathbf{r}, t)}$ is governed by the coupled equations
\begin{gather}
  \frac{\partial n_i}{\partial t} + \frac{\hbar}{m} \nabla \cdot (n_i \nabla \phi_i) = 0 \,,  \label{S16}
  \\
  \frac{\partial \phi_i}{\partial t} = -\frac{1}{\hbar} \frac{\delta E}{\delta n_i} \,,  \label{S17}
\end{gather}
where Eq.~(\ref{S16}) is the continuity equation, and Eq.~(\ref{S17}) is the generalization of the Josephson relation~\cite{PethickBook}.

For small-amplitude density variation, Eq.~(\ref{S16}) and (\ref{S17}) reduce to the linearized hydrodynamic equation
\begin{align}
  \frac{\partial^2}{\partial t^2} \delta n_i  = -\frac{\hbar^2}{4m^2} \nabla^2 \big(\nabla^2\delta n_i \big) + \frac{n_i}{m} \sum_{j=1,2} \frac{\partial \mu_i }{\partial n_j}\delta n_j + \alpha_i n_i \frac{q^2}{m} \left(e^{i(\mathbf{q\cdot r}-\omega t)}+ \textrm{c.c.}\right)e^{\eta t} \,,  \label{S18}
\end{align}
whose solution can be written as
\begin{align}
  \delta n_i(\mathbf{r},t) = \sum_{j=1,2} \alpha_j \left[ \chi_{ij}(q,\omega) e^{i(\mathbf{q\cdot r}-\omega t) } + \textrm{c.c.}\right] e^{\eta t} \,.  \label{S19}
\end{align}
Here $\chi_{ij}(q,\omega)$ is the species-resolved density response function
\begin{gather}
  \chi_{ii} (q,\omega) = \frac{2\epsilon_q n_i}{m (c_+^2-c_-^2)} \sum_{\ell = \pm} \ell \frac{\gamma_{3-i}n_{3-i}-mc_\ell^2 }{\hbar^2 \big[(\omega+\mathrm{i}\eta)^2 - \omega_{q,\ell}^2\big]} \, ,  \label{S20}
  \\
  \chi_{12} (q,\omega)=\chi_{21}(q,\omega)= -\frac{2\epsilon_q\gamma_{12}n_1n_2}{m(c_+^2-c_-^2)} \sum_{\ell = \pm} \frac{\ell }{\hbar^2 \big[(\omega+\mathrm{i}\eta)^2 - \omega_{q,\ell}^2\big]} \, ,   \label{S21}
\end{gather}
with $\epsilon_q = \hbar^2q^2/2m$, and the energy of elementary excitations $\hbar\omega_{q,\ell}=\sqrt{\epsilon_q(\epsilon_q+2mc_\ell^2)}$. The sound velocities $c_\pm$ are discussed in the following subsection.

\subsection{2.2 \quad Sound Velocity}

The long-wavelength excitations consist of two phonon modes: the lower-branch ($\ell=-$) corresponds to the in-phase density oscillation, and the upper-branch ($\ell=+$) corresponds to the out-of-phase oscillation.
The sound velocities of the two branches are given by [Eq.~(12) in the main text]
\begin{align}
  c_\pm = \sqrt{ \frac{1}{2m} \Big[\gamma_1 n_1 + \gamma_2 n_2 \pm \sqrt{ ( \gamma_1 n_1 - \gamma_2 n_2 )^2 + 4 \gamma_{12}^2 n_1n_2} \Big] } \,,
  \label{Ssound}
\end{align}
where $\gamma_1$, $\gamma_2$ and $\gamma_{12}$ can be further expressed in terms of densities, as derived in Eqs.~(\ref{gamma1})-(\ref{gamma12}).
For the self-bound liquid ($n_i=n_i^{(0)}$), we find 
\begin{gather}
  c_+^{(0)} = \sqrt{\frac{gn^{(0)}}{m}} \left[ 1 + \frac{8}{\sqrt{\pi}} \bigg(\lambda+ \frac{1}{\lambda} -\frac45 \bigg) \sqrt{n^{(0)} a^3} \right],
\\
  c_-^{(0)} = 4\sqrt{\frac{gn^{(0)}}{5\sqrt{\pi}m}\sqrt{n^{(0)} a^3}}. \label{S24}
\end{gather}
Our expression for $c_-^{(0)}$ is consistent with the previous result based on the Beliaev theory~\cite{Yin,Yin2} to the order $|\delta \tilde g|^{3/2}$. In the spinodal region ($\gamma_1\gamma_2<\gamma_{12}^2$), the lower-branch sound velocity $c_-$ becomes imaginary, indicating a dynamical instability.

We note that, in our hydrodynamic theory, the Andreev-Bashkin effect is not taken into account, which would results in a minor correction to the sound velocity~\cite{ABTheory1,ABTheory2}. In general, such a correction mainly affects the out-of-phase phonon mode, and its magnitude is much smaller than that of the Lee-Huang-Yang term~\cite{ABTheory2}.

\subsection{2.3 \quad Structure Factor and Correlation Lengths}

The pair-distribution function $\mathcal{D}_{ij}(r)\equiv\frac{1}{n_in_j}\langle \hat\Psi_i^\dagger(\mathbf{r}) \hat \Psi_j^\dagger(\mathbf{0}) \hat \Psi_j(\mathbf{0})\hat \Psi_i(\mathbf{r}) \rangle $ measures the relative probability of finding two particles of a given species at distance $r$, where $\hat \Psi_{i}(\mathbf{r})$ is the field operator and $r=|\mathbf{r}|$.
$\mathcal{D}_{ij}(r)$  is related to the static structure factor $S_{ij}(q)$ through the relation~\cite{FeenbergBook,LevSandroBook}
\begin{align}
  \mathcal{D}_{ij}(r) \, & = \, 1 + \frac{1}{\sqrt{n_in_j}} \int \frac{\mathrm{d}\mathbf{q}}{(2\pi)^3} \left[ S_{ij}(q) -\delta_{ij} \right]e^{-i\mathbf{q\cdot r}}
  \nonumber \\
   & = \, 1 + \frac{1}{\sqrt{n_in_j}}  \int_0^\infty \mathrm{d}q\, \left[ S_{ij}(q) -\delta_{ij} \right] \frac{q }{2\pi^2 r} \sin qr  \, .  \label{S27}
\end{align}
and $S_{ij}(q)$ can be constructed from the density response function $\chi_{ij}(q,\omega)$ by applying the fluctuation-dissipation theorem~\cite{LandauBook}, which at zero temperature takes the form
\begin{align}
S_ {ij} (q) = \frac{\hbar}{\pi\sqrt{n_in_j}} \int_0^\infty d\omega \, \mathrm{Im} \chi_{ij} (q,\omega) \,.
\end{align}
By inserting Eqs.~(\ref{S20}) and (\ref{S21}) for $\chi_{ij}$, and using the identity $\mathrm{Im} \frac{1}{x+i0^+} = -i\pi \delta (x)$, we obtain
\begin{gather}
  S_{ii} (q) \, = \,  \frac{\epsilon_q}{m(c_+^2-c_-^2)}\sum_{\ell=\pm} \ell \frac{mc_\ell^2-\gamma_{3-i}n_{3-i}}{\hbar\omega_{q,\ell}} \, , \label{S25}
  \\
  S_{12}(q) \, = \, S_{21}(q) \,=\, \frac{\epsilon_q}{m(c_+^2-c_-^2)} \sum_{\ell=\pm} \ell \frac{\gamma_{12}\sqrt{n_1n_2}}{\hbar\omega_{q,\ell}} \, . \label{S26}
\end{gather}

In the long-wavelength limit, $S_{ij}(q)$ is a linear function of $q$, which can be explicitly written as
\begin{align}
  S_{ij}(q\rightarrow 0) = \frac{q\xi_{ij}}{\sqrt{2}} \,.  \label{S28}
\end{align}
Such a small-$q$ behavior is actually a general feature of interacting superfluids at zero temperature~\cite{LevSandroBook,FeenbergBook}. It implies, at large distance $r$, the asymptotic form of $\mathcal{D}_{ij}$ reads
\begin{align}
  \mathcal{D}_{ij}(r\rightarrow \infty) = 1- \frac{\xi_{ij}}{\sqrt{2n_in_j} \pi^2 r^4} \,.  \label{S29}
\end{align}
In deriving this result, we have used the following expansion for $r\rightarrow \infty$~\cite{LighthillBook}
\begin{align}
  \int_0^\infty \mathrm{d} q \, f(q) \sin qr \, \sim \, \frac{f(0)}{r} - \frac{f''(0)}{r^3} + \mathcal{O}\big(r^{-5}\big) \, ,
\end{align}
which holds when the function $f(q)$ and all its derivatives are well-behaved.

Equation (\ref{S28}) can be viewed as a definition of the correlation lengths $\xi_{ij}$. Therein, the prefactor $\frac{1}{\sqrt{2}}$ is introduced to be consistent with the usual convention in a single-component Bose gas~\cite{LevSandroBook}.  By expanding the right-hand side of Eqs.~(\ref{S25}) and (\ref{S26}) to first order in $q$, we find
\begin{gather}
  \xi_{ii} \, = \, \frac{\hbar}{\sqrt{2}m(c_++c_-)} \Big( 1 + \frac{\gamma_{3-i}n_{3-i}}{mc_+c_-} \Big) \,, \label{S30}  \\
  \xi_{12} \, = \, \xi_{21} \,=\, -\frac{\hbar}{\sqrt{2}} \frac{\gamma_{12} \sqrt{n_1n_2}}{m^2(c_++c_-)c_+c_-} \, . \label{S31}
\end{gather}

\section{III. \quad Critical Behavior of Liquid-Gas Transition}

In this section, we present details in characterizing the critical behaviors, including divergent susceptibility, phonon-mode softening, and enhanced correlation lengths. These quantities are derived using the hydrodynamic approach presented in the preceding section.

\subsection{3.1 \quad Divergent Susceptibility}

In the static and long-wavelength limit, the density response function reduces to the thermodynamic susceptibility $\chi_{ij}^0\equiv \Big( \frac{\partial n_i}{\partial \mu_j} \Big)_{\mu_{3-j}}$, which measures the uniform density variation induced by a small shift of the chemical potentials. Explicitly, using the properties of Jacobians, we find
\begin{align}
  \chi_{ij}^0  \,=\, \left(\frac{\partial n_i}{\partial \mu_j}\right)_{\mu_{3-j}}
  = \, \frac{\partial (n_i,\mu_{3-j})}{\partial (n_j, n_{3-j})} \frac{\partial (n_j,n_{3-j})}{\partial (\mu_j, \mu_{3-j})} \,
  = \, \begin{cases}
    \dfrac{\gamma_{3-i}}{\gamma_1\gamma_2-\gamma_{12}^2},  & i=j \\
    \dfrac{-\gamma_{ij}}{\gamma_1\gamma_2-\gamma_{12}^2},  & i\neq j \\
  \end{cases} \,.
\end{align}
The same result can be deduced from Eqs.~(S29) and (S30), by imposing $\omega=0$ and $q\rightarrow 0$ therein.

At the critical points where the liquid-gas coexistence lines meet the diffusive spinodal (with $\gamma_1\gamma_2=\gamma_{12}^2$), all the static susceptibilities become divergent.


\subsection{3.2 \quad Phonon-Mode Softening }

At either critical point, the sound velocity vanishes along with the minority density. But it is the linear density dependence of the sound velocity that stands out as a critical phenomenon. Specifically,
for $n_1$ fixed at $n_1^\textsc{c}$, a straightforward small-$n_2$ expansion of $c_-$ gives
\begin{align}
  c_- = n_2 \sqrt{\frac{1}{m}\frac{\partial}{\partial n_2}\Big(\gamma_2-\frac{\gamma_{12}^2}{\gamma_1}\Big)} \Bigg|_{(n_1,n_2)=(n_1^\textsc{c},0)} = \frac{5\sqrt{\lambda}}{2\sqrt{2}} c_-^{(0)} \tilde n_2 \, ,
  \label{Sc1}
\end{align}
where $\tilde n_2=n_2/n_2^{(0)}$, and in the second equality we have used Eq.~(\ref{S24}). Such a linear dependence only occurs at the critical point. Indeed, for any fixed $n_1>n_1^\textsc{c}$, the expansion yields
\begin{align}
  c_- = 2\sqrt{1+\lambda}\,c_-^{(0)} \sqrt{\tilde n_2 \Big(\!\sqrt{\frac{n_1}{n_1^\textsc{c}}}-1\Big)} \,. \label{Sc2}
\end{align}
The clear distinction between Eq.~(\ref{Sc1}) and (\ref{Sc2}) is illustrated in Fig.~3(a) of the main text.

\subsection{3.3 \quad Enhanced Correlation Lengths}

From Eqs.~(\ref{S30}) and (\ref{S31}), it is also straightforward to derive the critical correlation lengths.
For $n_1$ fixed at $n_1^\textsc{c}$, the correlation lengths in the small minority-density limit behave like
\begin{align}
  &\xi_{11} \xrightarrow{n2\rightarrow0}\, \xi_1^\textrm{single} \Big( 1+ \frac{1}{\sqrt{|\delta \tilde g|}}\Big) \,,
  \\
  &\xi_{12} \xrightarrow{n2\rightarrow0} \frac{\xi^{(0)}}{\sqrt{3(1+\lambda)\tilde n_2}} \rightarrow \infty \,,
  \\
  &\xi_{22} \xrightarrow{n2\rightarrow0}  \frac{4\,\xi^{(0)}}{5\sqrt{3\lambda}\,\tilde n_2} \rightarrow \infty  \,.
\end{align}
By contrast, for any fixed $n_1> n_1^\textsc{c}$, we find following results in the small minority-density limit
\begin{align}
  &\xi_{11} \xrightarrow{n2\rightarrow0} \, \xi_1^\textrm{single} \,,
  \\
  &\xi_{12} \xrightarrow{n2\rightarrow0}  \frac{5\sqrt{\lambda}}{4\sqrt{6}(1+\lambda)} \frac{\xi^{(0)}}{\sqrt{\tfrac{n_1}{n_1^\textsc{c}}\Big(\!\sqrt{\tfrac{n_1}{n_1^\textsc{c}}}-1\Big)}} \,,
  \\
  &\xi_{22} \xrightarrow{n2\rightarrow0} \frac{\xi_0}{\sqrt{6(1+\lambda)\Big(\!\sqrt{\tfrac{n_1}{n_1^\textsc{c}}}-1\Big)\, \tilde n_2}} \rightarrow \infty
\end{align}
These are illustrated in Fig.~3(b)(c)(d) of the main text.